\documentclass[12pt]{iopart}
\begin{document}
\title[Hamilton's equations for a fluid 
membrane]
{Hamilton's equations for a fluid membrane}
\author{R  Capovilla\dag ,
J Guven\ddag and E Rojas\S}
\address{\dag\
Departamento de F\'{\i}sica,
 Centro de Investigaci\'on
y de Estudios Avanzados del IPN,
Apdo Postal 14-740, 07000 M\'exico,
D. F.,
MEXICO}
\address{\ddag\
Instituto de Ciencias Nucleares,
 Universidad Nacional Aut\'onoma de M\'exico,
 Apdo. Postal 70-543, 04510 M\'exico, DF, 
MEXICO}
\address{\S\ Facultad de F\'{\i}sica e 
Inteligencia
Artificial, Universidad Veracruzana, 91000 
Xalapa, Veracruz,
MEXICO}

\begin{abstract}
 Consider a homogenous fluid membrane described by the Helfrich-Canham  
energy, quadratic in the mean curvature of the membrane surface. 
The shape equation that determines equilibrium configurations
is fourth order in derivatives 
and cubic in the mean curvature.
We introduce a Hamiltonian formulation of this equation which dismantles 
it into a set of coupled 
first order equations. This involves  interpreting 
the Helfrich-Canham energy 
as an action; 
equilibrium surfaces are generated by the evolution of space curves.
Two features 
complicate the implementation of a Hamiltonian framework: ($i$) The action 
involves second derivatives. This  
requires treating the velocity as a phase space variable and the
introduction of its conjugate momentum. The canonical 
Hamiltonian is constructed on this phase space.
($ii$) The action possesses a local symmetry -- reparametrization 
invariance. 
The two labels we use to parametrize points on the surface are 
themselves physically irrelevant.
This symmetry implies primary constraints, one for each label, 
that need to be implemented within the Hamiltonian. The 
two lagrange multipliers 
associated with these constraints are identified as 
the components of the acceleration tangential to the surface. 
The conservation of the primary constraints 
imply two secondary constraints, fixing the 
tangential components of the momentum conjugate to the position. 
Hamilton's equations
are derived and the appropriate initial conditions on the  
phase space variables are identified. 
Finally, it is shown how the shape equation
can be  reconstructed from these equations. 
\end{abstract}
\today


\maketitle

\section{ Introduction}

A biological membrane consists, in a first approximation, of a bilayer of 
phospholipid molecules in water \cite{handbook,boal}. 
The thickness of the membrane is determined by the 
length of the hydrocarbon tails $\sim 50\, nm$, whereas its overall size can reach 100
$\mu m$ --  it is thus sensible to describe it on mesoscopic scales 
as a two-dimensional surface. Remarkably, many of the physical 
properties of the membrane are completely described by the geometrical degrees of freedom 
associated with this surface. The language that is 
most appropriate to  describe the membrane is differential geometry. 

The phospholipid  molecules move freely within the membrane
so that the membrane behaves as a two-dimensional fluid. In a geometrical model,
this feature is implemented by associating  
an energy, invariant under surface reparametrizations,
with each possible shape of the membrane.
One should think of reparametrization in its active sense, as
the freedom of a point to move about on the surface, rather than as a  
change of coordinates. 

In the early seventies, it was recognized that the leading term in
the energy is proportional to the cost of bending the
membrane  -- quadratic in the mean curvature 
\cite{canham,helfrich,evans}.
This geometric invariant is known 
in the mathematical literature as the Willmore functional \cite{Willmore}.
The geometry of  a fluid membrane will generally also be subject to constraints:
its area and (if closed) its volume are constant.
Moreover,  the bilayer structure is captured 
by the area difference between the layers, 
proportional to the integrated mean 
curvature. Its value is constrained  in the  minimal curvature model for 
fluid membranes \cite{svetina89}.  
This is the model we will focus our attention on. 
Other curvature models have superseded this 
simple model (see {\it e.g.} \cite{seifert} for a review). In particular,
it is worth mentioning 
the ADE model, with a non-local constraint involving the square 
of the area difference \cite{ADE1,ADE2,ADE3}. Our
considerations  can be extended to this and other geometrical models.

We are interested, in particular, in the equilibrium 
shapes of a closed fluid vesicle minimizing the Helfrich-Canham energy,
defined below by Eq.(\ref{eq:model}), which 
describes the minimum curvature model. 
Typically, the addition of 
constraints, will increase the number of equilibrium solutions.
The vanishing of the first variation of the 
energy is expresssed in a single `shape' equation
\cite{ZH1,ZH2,stress}. It is a fourth
order non-linear PDE. Beginning in the midseventies, with the 
work of Deuling and Helfrich on axisymmetric configurations \cite{DH}, 
considerable 
effort and ingenuity has been dedicated to characterizing these shapes.
For axisymmetric configurations, it is relatively straightforward to 
solve the shape equation: the symmetry reduces it to a 3rd order non-linear 
ODE.  Even in this simple case, however, solving this equation 
involves numerical approximations \cite{seifert}.
For more general configurations, analytical techniques have been as good as useless
outside of perturbation theory. Rather than attempt to solve the shape equation itself,
one has relied on a combination of computational techniques, among them 
Monte-Carlo simulations and dynamical triangulations to 
minimize the energy (see {\it e.g.} \cite{Gompper,Bowick}).  

In principle, how does one go about  
solving the shape equation when axisymmetry is relaxed? The method used to 
solve  the axisymmetric shape equation numerically suggests a 
strategy. 
What one does is decompose the higher order shape equation into
a set of coupled first order equations. 
Initial conditions for these 
equations are specified, that will 
need to be tuned  to obtain a geometry which closes smoothly. 
In practice, modulo the step involving tuning, 
what one does is  formulate the problem  in Hamiltonian terms.
In this paper, we provide a novel formulation  of the 
shape equation, in all its glory, as a Hamiltonian sytem.
The basic idea is borrowed from the Hamiltonian formulation of 
General Relativity, and more specifically, from
the  recent Hamiltonian formulation of the dynamics of relativistic extended 
objects \cite{hamrel}.  Take any curve on the membrane surface.
The surface is then generated by the
evolution of the curve along some direction with appropriate initial data.
The evolution is determined by a Hamiltonian that is derived from
the Helfrich-Canham energy, reinterpreted as an action for the curve.
To do this in one step is daunting technically. 
The reader may be interested in first working through the details in  a
simplified setting. The axisymmetric case has 
been considered in \cite{hamaxi},
where  the shape equation is formulated  as
a Hamiltonian system describing the motion of a particle, rather than a 
curve, in 
two dimensions. While the canonical structure itself is illustrated clearly 
in this restricted setting, it is too simple to capture the geometry.

The paper is organized as follows.
In Sect. 2, we introduce the geometry for the surface that
describes the fluid vesicle, and the Helfrich-Canham energy.
Next, we consider an arbitrary parametrized closed curve on the surface,
and decompose the surface geometry along a tangential basis adapted to the 
curve: one vector is tangent to the curve, the other, not necessarily 
orthogonal to it, is parameterized by a second parameter $t$.
The latter tangent vector  acquires an 
interpretation as the velocity of the curve,
parameterized by the time $t$.
By implementing this decomposition of the surface geometry 
within the Helfrich-Canham
energy, it can be written as an action describing the evolution of the curve.
The corresponding Lagrangian
depends not only on the
position of the curve and its velocity, but also on its 
second derivative with respect to the parameter $t$: its acceleration.
A second key feature is the
presence of a local symmetry of the action, reparametrization invariance.
Before looking at lipid vesicles,
in Sect. 3 we consider the simpler, but still non-trivial,
case of a soap bubble. The  energy
is proportional to the area of the bubble with a constraint on the 
enclosed volume. The action does not involve second derivatives and 
we can address the problems associated with  reparametrization invariance without the added 
complication of higher derivatives. 
In Sect. 4 we return to fluid vesicles described by the
Helfrich-Canham energy, and we identify 
the phase space variables which are appropriate when the 
Lagrangian involves the acceleration; it becomes  
necessary to introduce canonical momenta not only conjugate to the
position of the curve, but also to its velocity. 
We show how these momenta can be identified 
by examining the response of the energy to deformations of this curve. 
The momentum conjugate to position will be shown to be  
related directly to the conserved stress introduced by two of
us in \cite{stress}. Not surprisingly, the stress plays a role in the 
Hamiltonian formulation; stresses get transmitted across closed curves \cite{MDG}.
In Sect. 5, we construct the canonical Hamiltonian
in terms of a Legendre transformation with respect to both
position and velocity. This Hamiltonian will generate the evolution of the 
curve in terms of the phase space variables.
At a Hamiltonian  level, the existence
of a local symmetry, in our case reparametrization invariance, manifests 
itself in the appearance of constraints on the phase space variables
at fixed values of the evolution parameter. We find two primary constraints,
that need to be added to the canonical Hamiltonian to generate evolution,
as well as two secondary constraints. 
These four constraints are conserved in
time. In particular, one of the secondary constraints is the
vanishing of the canonical Hamiltonian -- the hallmark of reparametrization
invariance. The secondary constraints are the generators of 
reparametrizations.
In Sect. 6, we derive the Hamilton equations for this system. 
Appropriate initial conditions
on the  phase space variables are discussed. Reparametrization  
invariance manifests itself in the appearance of two arbitrary functions 
in these equations. 
At the end
of the day the shape equation has been dismantled into its Hamiltonian building blocks:
four constraints at fixed value of time and four first order, in 
time, evolution equations. In Sect. 7,
we show how the shape equation
can be reconstructed using these blocks. 
We conclude in Sect. 8 with some final remarks. Various technical
details are collected in a set of Appendices.

\section{ Helfrich-Canham  energy}

We model a fluid membrane as  a two-dimensional surface $\Sigma$,
defined locally by the embedding
\begin{equation}
{\bf x} = {\bf X} (\xi^a )\,,
\end{equation}  
where ${\bf x} = x^i$ are local coordinate in space, $\xi^a $  local
coordinates on the surface $\Sigma$ ($i,j, \dots = 1,2,3; a,b, \dots =
1,2$), and ${\bf X} = X^i$  the
embedding, or shape, functions. We denote by ${\bf e}_a = \partial {\bf X}
/ \partial \xi^a$ the two tangent vectors to the surface.
The metric induced on $\Sigma$ is given by
their inner product, $g_{ab} =
{\bf e}_a \cdot {\bf e}_b $, and its inverse is $g^{ab}$;
indices are lowered and raised with $g_{ab}$ and $g^{ab}$, respectively.
The area of the surface $\Sigma$ is $ A = \int dA = \int d^2 \xi \sqrt{g}$, 
with $g =$ det$g_{ab}$.
The unit normal ${\bf n}$ to $\Sigma$ is defined
 implicitly by ${\bf e}_a \cdot {\bf n} = 0$, ${\bf n}^2 = 1$. 
The extrinsic curvature tensor is $K_{ab} = - {\bf n} \cdot
\partial_a \partial_b {\bf X}$, and the mean curvature is
$K = g^{ab} K_{ab}$.
In terms of the principal curvatures, $\{c_1 , c_2 \}$, we have
$K = c_1 + c_2$.
The intrinsic scalar curvature can be given in terms of the extrinsic
curvature via the
Gauss-Codazzi equation
as ${\cal R} = K^2 - K^{ab} K_{ab}$; it is twice the Gaussian 
curvature, so that in terms of the principal curvatures 
 ${\cal R} = 2 c_1 c_2$.   

The shape of a fluid membrane is  obtained by minimizing the
Helfrich-Canham  energy, quadratic in the extrinsic curvature.
In the bilayer coupling model, the membrane is subject to various global 
constraints: the enclosed volume, the 
area, as well as the integrated mean curvature 
(the area difference between the layers) are fixed. 
One then  searches for stationary points of the Helfrich-Canham energy
$F[{\bf X}]$ defined by 
\begin{equation}
F [{\bf X}] = {\kappa \over 2} \int dA \; K^2 + \beta \int dA \; K
+ \sigma A - {\sf P} V\,. 
\label{eq:model}
\end{equation}
The constant $\kappa$ is the bending rigidity;
$\beta$, $\sigma$ and ${\sf P}$ are Lagrange multipliers
enforcing the constraints of constant integrated mean curvature , constant area and
constant enclosed volume $V$, respectively.
The constant $\sigma$ is the bare surface tension, and ${\sf P}$ is 
the osmotic pressure. Of course, the inclusion of the volume term requires
that the surface be closed.
The volume can be written
as a surface integral:
\begin{equation}
V = {1 \over 3} \int dA \; {\bf n} \cdot {\bf X}\,.
\label{eq:vol}
\end{equation}   
Despite appearances, this expression for $V$ is translationally invariant. 
This is because 
$\int dA \, {\bf n}\cdot {\bf a} =0$ for any constant vector ${\bf a}$
on any closed surface.  
We have not included in the  energy a term corresponding to the 
Gaussian 
bending, $F_G [{\bf X}] = \kappa_G \int dA {\cal R}$, with $\kappa_G$ 
the Gaussian bending rigidity, since for a two-dimensional surface, by the 
Gauss-Bonnet theorem,  it is a topological invariant, and so it does not
contribute to the determination of equilibrium configurations.

The vanishing of the first variation of the  energy (\ref{eq:model}),
with respect to infinitesimal variations of the shape functions
${\bf X}  \to
{\bf X} + \delta {\bf X}$, gives the shape equation
\cite{ZH1,ZH2,stress,Auxil}
\begin{equation}
\kappa \left[ -  \nabla^2 K - {K \over 2} ( K^2 - 2 {\cal R} ) \right]
+ \beta {\cal R} + \sigma K - {\sf P} = 0\,,
\label{eq:shape}
\end{equation}
where $\nabla^2 = g^{ab} \nabla_a \nabla_b$ denotes the surface Laplacian, and $\nabla_a$
the surface covariant derivative. 
This fourth order non-linear PDE determines the
equilibrium configurations of fluid vesicles.
Note that there is only one equilibrium condition. Naively varying ${\bf X}$,
one would have expected three, since there are three independent 
deformations. However, 
reparametrization invariance tells us that two of the equations, 
corresponding to
tangential deformations, must vanish identically \cite{Second}. The
only physical deformations that enter in the determination of equilibrium 
are those normal to the surface. 

Our basic proposal in this paper is to consider the
surface $\Sigma$ as the evolution of some closed
curve ${\cal C}$. If the curve is parametrized by 
$u$, and its evolution by $t$, the surface generated by 
${\cal C}$ will be given by ${\bf x}={\bf X}(u,t)$.
For any fixed $t$, the tangent vector to the curve is ${\bf X}' = 
\partial_u {\bf X} =
\partial {\bf X} / \partial u$.
The infinitesimal
arclength along the curve is $ds = h^{1/2} du$, where 
\begin{equation}
h = {\bf X}'\cdot {\bf X}'
\label{eq:hdef}
\end{equation} is  
the one-dimensional metric along ${\cal C}$; 
the unit tangent vector is ${\bf 
t} = h^{-1/2} {\bf X}'$.
The curve ${\cal C}$ evolves along the surface vector field
$\dot {\bf X} = \partial_t {\bf X} =  
\partial {\bf X} / \partial t$. This vector will not
generally be orthogonal to ${\cal C}$. 
Let ${\bf l}$ denote the unit vector normal to the curve on the surface 
${\bf l}\cdot {\bf X}'=0$.
We expand $\dot {\bf X}$ with respect to the vectors ${\bf l}$ and ${\bf X}'$:
\begin{equation}
\dot{\bf X} = N {\bf l} + N_\parallel {\bf X}'\,;
\label{eq:velo}
\end{equation}
in analogy with the terminology used in general relativity, we will refer to 
the components $ N, N_\parallel$ with respect to this basis as 
the lapse and shift functions, respectively.
In terms of ${\bf X}'$ and $\dot {\bf X}$, they are given by
\begin{equation}
N^2 = \dot{\bf X}^2 -  {(\dot {\bf X} \cdot {\bf X}')^2\over h}\,, \quad
N_\parallel = {(\dot {\bf X} \cdot {\bf X}')\over h }\,.
\label{eq:NN}
\end{equation}
We now decompose the geometry of the surface $\Sigma$ 
along the basis of tangent vectors
adapted to the evolution of the curve,
$\{\dot{\bf X}, {\bf X}' \}$. 
For the induced metric we have
\begin{equation}
g_{ab} 
=
\left(\matrix{ N^2 +  h N_{||}^2 & h N_{||}\cr
                            h N_{||}          & h\cr}\right)
\,;
\end{equation}
and for its inverse it follows that
\begin{equation}
g^{ab} = N^{-2} \left(\matrix{ 1   & - N_{||}\cr
                             -N_{||}     &  N^2 h^{-1} + N_{||}^2\cr}\right)
\,.
\end{equation}
Note that the component $g^{uu}$ can be written also as $g^{uu} = \dot{\bf 
X}^2 / h$. The determinant of the induced metric $g_{ab}$ takes the  
simple form \begin{equation} g = N^2 h \,.
\label{eq:det}
\end{equation}
The vectors ${\bf t}, {\bf l}$,
together
with the normal to the surface ${\bf n}$ form a spatial basis adapted
to the curve. We choose an orientation so that
${\bf n} = {\bf t} \times {\bf l}$. In terms of ${\bf X}'$ and $\dot{\bf X}$,
\begin{equation}
{\bf n}= {1 \over N \sqrt{h}} \; {\bf X}'\times \dot {\bf X}\,.
\label{eq:ndefx}
\end{equation}

The decomposition of the extrinsic curvature along the basis $\{ \dot{\bf 
X}, {\bf X}' \}$ is given by 
\begin{equation}
K_{ab} = - \left(\matrix{  {\bf n} \cdot \ddot{\bf X}  & {\bf n} \cdot
\dot{\bf X}' \cr  {\bf n} 
\cdot \dot{\bf X}'     &   {\bf n} \cdot {\bf X}''     
\cr}\right) \,,
\end{equation}
which depends on the acceleration of the curve $\ddot{\bf X}$ projected
along the normal to the surface.
For the mean curvature it follows that
\begin{equation}
K = g^{ab} K_{ab} = {1 \over N^2} \left( - {\bf n} \cdot \ddot{\bf X}  + J 
\right)\,, \label{eq:kpro}
\end{equation}
where we have have isolated the part that 
does not depend on the acceleration in the
quantity 
\begin{equation}
J = 2 N_\parallel \, ( {\bf n} \cdot  \dot{\bf X}') - h^{-1} \, ( 
\dot{\bf X}^2 ) \,  ({\bf n} \cdot {\bf X}'') \,.
\label{eq:J}
\end{equation}

To make contact with the analysis in \cite{hamaxi} for axisymmetric
configurations, recall 
that an axisymmetric configuration 
can be described by cylindrical coordinates $\{ R(t), Z(t), \phi \}$. 
The curve ${\cal C}$ is parametrized
by the angle $\phi$; the tangent vector is ${\bf X}' = \partial {\bf X} / 
\partial \phi$, and the one-dimensional metric is $h = R^2$.
The velocity $\dot{\bf X}$ is along a meridian, with $N_\parallel = 0$, and 
$N = \sqrt{ \dot{R}^2
+ \dot{Z}^2}$. Moreover,  in this case,
${\bf n} \cdot \dot{\bf X}' = 0$, ${\bf n} \cdot {\bf X}''
= - N^{-1} R \dot{Z}$, and ${\bf n} \cdot \ddot{\bf X} = N^{1/2}
( \dot{Z} \ddot{R} - \dot{R} \ddot{Z} )$; it follows that
the quantity $J$ takes the simple form $J = R^{-1} N \dot{Z}$, so that
the mean curvature is
$ K = R^{-1} N^{-3/2} [ R ( \dot{R} \ddot{Z} - \dot{Z} \ddot{R} ) + N^2 
\dot{Z} ]$. 

We now cast the energy $F[{\bf X}]$ given by 
(\ref{eq:model}) 
as an action decribing the dynamics in time $t$ of a curve
parametrized by $u$, ${\bf X}(t,u)$:
\begin{equation}
F [{\bf X}] = \int dt \,
L [{\bf X}, \dot{\bf X} , \ddot{\bf X} ]\,,
\label{eq:action}
\end{equation}
where the Lagrangian functional is given by
\begin{equation}
\fl
L [{\bf X}, \dot{\bf X} , \ddot{\bf X} ] 
= \oint du \; N \; \sqrt{h} \left[  
{\kappa  \over 2 N^4} \left( - {\bf n} \cdot 
\ddot{\bf X}  + J \right)^2
+ {\beta \over N^2 } \left( - {\bf n} \cdot \ddot{\bf X}  + J \right)
+ \sigma   
-  {{\sf P}\over 3}   {\bf n} \cdot {\bf X} \right]\,.
\label{eq:lagrange}
\end{equation}
Note the dependence of the Lagrangian on time 
derivatives of the field variables ${\bf X}$:
it is {\it quadratic} in $\ddot {\bf X}$, the acceleration of 
the curve, and it is not  
possible to eliminate this dependence using integration by parts 
within the action. It depends on the velocity $\dot{\bf X}$ through its 
appearence in $N$,  $J$ and ${\bf n}$.
In all but the last term, 
the position ${\bf X}$ appears only through its derivatives 
with respect to the parameter $u$ along the curve, 
${\bf X}'$ and ${\bf X}''$; note that there is an implicit dependence
on ${\bf X}$ of $N$ and ${\bf n}$.

The Hamiltonian formulation of the
dynamical system defined by the action (\ref{eq:action}) is somewhat involved. 
As an aid to establishing our bearings, it will be useful to first understand 
the Hamiltonian formulation of a simpler
system, without higher derivatives to contend with.

\section{ Soap bubble}

In this section we will provide a Hamiltonian formulation of the equilibrium
of a soap bubble, in which surface tension is pitted against the 
internal pressure.
Here the energy is simply proportional to the area of the
bubble, subject to a constraint fixing the enclosed volume. The 
energy involves no higher  than the first derivatives of ${\bf X}$
with respect to the parameter $t$.
This system  illustrates, at the Hamiltonian
level, the consequences of the presence of a local symmetry
-- reparametrization invariance -- namely the presence of constraints
on the phase space variables. The 
reader should be warned, however, that the canonical structure is
very  different from the one presented below for a lipid vesicle.

The  energy is
\begin{equation}
F [{\bf X}] = \sigma A - {\sf P} V \,,
\label{eq:soap}
\end{equation}
where $\sigma$ is the surface tension, and ${\sf P}$ the pressure excess 
inside the bubble. The equilibrium condition obtained from the vanishing of the first
variation of $F [{\bf X}]$ is given by
the classical Laplace-Young equation, of second order in derivatives of the
shape functions,
\begin{equation}
\sigma K - {\sf P} = 0\,.
\label{eq:LY}
\end{equation}
Equilibrium configurations are surfaces of constant mean curvature.

We consider the  energy  (\ref{eq:soap}) 
as an action, decribing the motion in the parameter $t$ of a curve
parametrized by $u$, ${\bf X}(t,u)$.
Decomposing  the energy with respect to the coordinates $\{ t , u \}$,
we have
\begin{equation} 
F [{\bf X}] =  \int dt \; L [ {\bf X},  \dot{\bf X} ]\,,
\label{eq:actionsoap}
\end{equation}
where
the Lagrangian functional is 
\begin{equation} L [ {\bf X}, \dot{\bf X}]
=  \oint du \; N \sqrt{h} \; \left(\sigma - {{\sf P} \over 3} \, {\bf n} 
\cdot {\bf X} \right) \,. 
\label{eq:l1}
\end{equation}
We emphasize that this Lagrangian  depends at most on the
velocities via $N$
and  ${\bf n}$ related respectively to $\dot {\bf X}$ by (\ref{eq:NN}) and 
(\ref{eq:ndefx}).  The factor of 
$\sqrt{h}$ appearing in (\ref{eq:l1}) ensures 
invariance of $L$ with respect to reparametrizations of the curve ${\cal C}$. 

Our first step towards a Hamiltonian formulation of the problem is the derivation 
of the canonical momentum ${\bf p}$ conjugate to ${\bf X}$. A direct
approach would be to define the canonical momentum ${\bf p}$ via the
functional derivative $
{\bf p} = \delta L / \delta \dot{\bf X}
$. However, with an eye towards the higher order generalization
for fluid vesicles, it is preferable to extract the canonical momentum
from the first variation of the action (\ref{eq:actionsoap}); 
we do this in some detail.
Under an infinitesimal deformation 
of the embedding functions ${\bf X} \to {\bf X} + \delta {\bf X}$,
the first variation of the action 
(\ref{eq:actionsoap}),
can be written as
\begin{equation}
\delta F [{\bf X}] = \int dA \, {\bf E} \cdot \delta {\bf X} + \oint
du \;{\bf p} \cdot \delta {\bf X}\,,
\label{eq:firsts}
\end{equation}
where ${\bf E}$ denotes the Euler-Lagrange derivative. This 
expression allows us to read off 
the canonical momentum ${\bf p}$ from the second term.
To carry out the first variation, 
we use the well known variational expressions (see {\it e.g.} \cite{Second})
\begin{eqnarray}
\delta A &=& \int dA \; {\bf e}^a \cdot \partial_a \delta {\bf X}\,,
\label{eq:va}
\\
\delta V &=& {1 \over 3} \int dA \; \left[ {\bf n} \cdot \delta {\bf X}
+ ({\bf n} \cdot {\bf X}) {\bf e}^a \cdot \partial_a \delta {\bf X}
- ({\bf e}^a \cdot {\bf X} ) {\bf n} \cdot \partial_a \delta {\bf X}    
\right]\,. 
\label{eq:vv}
\end{eqnarray}
Neither expression depends on the parametrization.
For the first variation of the energy (\ref{eq:actionsoap}) it follows that, 
integrating by parts it can be cast in the form (\ref{eq:firsts}),
\begin{equation}
\fl
\delta F = \int dA \left( - \sigma \nabla_a {\bf e}^a - {\sf P} {\bf n} 
\right) \cdot \delta {\bf X} + \int 
dA \nabla_a \left\{ \sigma {\bf e}^a  \cdot \delta {\bf X}   - 
{{\sf P} \over 3}  
\left[ ({\bf n} \cdot {\bf X} ) {\bf e}^a - ( {\bf e}^a \cdot {\bf X} ) {\bf 
n} \right]  \cdot \delta {\bf X} \right\}\,.
\end{equation}
Now we use
the Gauss-Weingarten equations for the surface $\Sigma$
\begin{eqnarray} 
\nabla_a {\bf e}_b &=& - K_{ab} {\bf n}\,, \label{eq:gw1} \\
\nabla_a {\bf n} &=& K_{ab} \; g^{bc} \; {\bf e}_c\,, \label{eq:gw2}
\end{eqnarray}
and Stokes' theorem as applied to a  surface vector field $V^a$ which 
states that, for a region of the surface enclosed by the curve ${\cal C}$,
\begin{equation}
\int dA \; \nabla_a V^a = \oint du \; \sqrt{h}  \; l_a \; V^a\,, 
\label{eq:stokes}
\end{equation}
where $l^a$ is the normal to ${\cal C}$ on the 
surface $\Sigma$ (note that ${\bf l} = l^a {\bf e}_ a$).
Therefore we obtain
\begin{equation}
\fl
\delta F = \int dA \left( \sigma K - {\sf P} \right) {\bf n}  \cdot \delta 
{\bf X}
+ \oint du  \sqrt{h} \left\{ \sigma {\bf l} \cdot \delta{\bf X}  - {{\sf P} 
\over 3}  
\left[ ({\bf n} \cdot {\bf X} ) {\bf l} - ( {\bf l} \cdot {\bf X} ) {\bf 
n} \right] \cdot \delta {\bf X} \right\} \,.
\label{eq:soapa}
\end{equation}
Comparison of (\ref{eq:firsts}) and (\ref{eq:soapa}) 
identifies  the Euler-Lagrange derivative as
${\bf E} = E {\bf n} $, where $E$ is given by the l.h.s. of 
(\ref{eq:LY}); the Euler-Lagrange  derivative is 
normal to the 
surface. For the canonical momentum we read off
\begin{eqnarray}
{\bf p} &=&
  \sqrt{h} \left\{ \sigma {\bf l}  - {{\sf P} \over 3}  
\left[ ({\bf n} \cdot {\bf X} ) {\bf l} - ( {\bf l} \cdot {\bf X} ) {\bf 
n} \right] \right\} \nonumber \\
&=&   \sigma \; \sqrt{h} \; {\bf l}  
- {{\sf P} \over 3} {\bf X} \times {\bf X}'\,. 
\label{eq:p1}
\end{eqnarray}
The momentum transforms as a density under reparametrizations
of the curve ${\cal C}$, as it should.
It is normal to the curve. In this sense, it differs in an essential way 
from $\dot{\bf X}$ which is not necessarily  normal. 
The contribution to ${\bf p}$ from area is proportional through $\sigma$ to the 
densitized unit vector ${\bf l}$. The addition to ${\bf p}$ due to the 
volume constraint is independent of $\dot{\bf X}$, which  
introduces a component normal to the surface. 

Note that neither $N$ nor $N_\parallel$ are canonical variables for the soap bubble: they cannot be expressed as
functionals of ${\bf X}$ and ${\bf p}$. They should not therefore appear within 
the canonical framework. For the same reason, 
the normal  ${\bf n}$ (unlike the tangent 
vectors ${\bf t}$ and ${\bf l}$) is  also not a  canonical variable,
and has no place within this framework in this model.  
For fluid membranes, on the other hand,  $N$, $N_\parallel$ as well as ${\bf n}$ will
be identified as functions of the canonical variables.

We have now identified the phase space variables appropriate
for a Hamiltonian description of the equilibrium configurations of a soap 
bubble: the position 
of a given curve ${\bf X}$ and its conjugate momentum ${\bf p}$, as given
by (\ref{eq:p1}).
The canonical Hamiltonian is obtained via a Legendre transformation
of the Lagrangian (\ref{eq:l1}) with respect to  $\dot{\bf X}$ as
\begin{equation}
H_0 [ {\bf p}, {\bf X}]
= \oint du \; {\bf p} \cdot \dot{\bf X} - L [ {\bf X}, \dot{\bf X} ]\,.
\end{equation}  
It is immediate to see that it vanishes,   
\begin{equation}
H_0 [ {\bf p}, {\bf X}]  = 0\,.
\end{equation}
Its vanishing is a direct consequence of the
presence of  
reparametrization invariance.
At first this might appear to indicate some inconsistency.
However, a local symmetry
such as reparametrization invariance also implies constraints. These constraints 
will generate the evolution. Let us examine these issues one at a time.
Reparametrization invariance renders  it   
impossible to invert for the velocity in terms of the momentum.
To see this explicitly one  considers the Hessian
\begin{equation}
H_{ij} = {\delta^2 L \over \delta \dot{X}^i \delta \dot{X}^j }
= {\sigma \sqrt{h} \over N} n_i n_j\,. 
\end{equation}  
$H_{ij}$ is degenerate: its determinant vanishes. This is a 
feature of any Hamiltonian system with local symmetries. 
It means that at any 
value of
the parameter $t$ the phase space variables are not all independent;
there are constraints. We use the definition of the momentum (\ref{eq:p1})
to identify the constraints as functions of the canonical variables:
${\bf p}$ is orthogonal to ${\cal C}$, and the function of the canonical 
variables 
$({\bf p}+ {\sf P} \,{\bf X} \times {\bf X}'/3)/\sigma \sqrt{h}$  is a unit vector.
We thus find that, at each value of $t$,
the phase space variables must satisfy the primary constraints:
\begin{eqnarray}
C_1 &=& \left( {\bf p} + {{\sf P} \over 3} {\bf X} \times {\bf X}'\right)^2
- \sigma^2 h = 0
\label{eq:sc1}\\
C_2 &=& {\bf p} \cdot {\bf X}' = 0\,.
\label{eq:sc2}
\end{eqnarray}
The only freedom ${\bf p}$ possesses independent of ${\bf X}$ is a direction.

The constraints possess a geometrical role in phase space as  generators of
infinitesimal reparametrizations. In particular, 
whereas $C_2$ generates 
reparametrizations of the curve ${\cal C}$ itself, $C_1$ generates motions 
off the curve ${\cal C}$ onto the surface. This can be shown  
by considering the Poisson bracket of the constraints with the
phase space variables, where,
for two arbitrary functions of the
phase space  variables $f$ and $g$,
the Poisson bracket is
\begin{equation}
\{ f , g \} = \oint  du \; \left[
{\delta f \over \delta {\bf X}}
\cdot {\delta g \over \delta {\bf p}}
- (f \leftrightarrow g ) \right]\,.
\label{eq:PBs}
\end{equation}
It should be mentioned that analogous constraints with 
${\sf P} = 0$ occur in the Hamiltonian formulation of a relativistic
bosonic string (see {\it e.g.} \cite{hamrel}).

The Hamiltonian that generates 
the motion of the curve ${\cal C}$ is proportional to the constraints
themselves,
\begin{equation}
H = \oint du \; ( \lambda_1 C_1 + \lambda_2 C_2 )\,,
\label{eq:hlc}
\end{equation}
where $\lambda_1 = \lambda_1 (u,t)$ and  $\lambda_2 = \lambda_2 (u,t)$  are 
two arbitrary Lagrange multiplier functions that enforce 
the constraints. The time derivative of a phase space function $f$ is
given by its Poisson bracket with this Hamiltonian,
\begin{equation}
\dot{f} = \{ f , H \}\,.
\end{equation}

In particular, it is important to note that the time derivative of the
constraints (\ref{eq:sc1}) and (\ref{eq:sc2}) does not generate 
any new constraints.

Hamilton's equations are given by (see the Appendices for useful formulas) 
\begin{eqnarray}
\fl
\dot{\bf X} = {\delta H \over \delta {\bf p  }}
&=& 2 \lambda_1 \left( {\bf p} + {{\sf P} \over 3} {\bf X} \times {\bf X}'
\right) + \lambda_2 {\bf X}'\,,
\label{eq:hs1} \\ \fl
\dot{\bf p} = - {\delta H \over \delta {\bf X}} &=& 
- {2 {\sf P} \over 3} \lambda_1 {\bf X}' \times \left( 
{\bf p} + {{\sf P} \over 3} {\bf X} \times {\bf X}' \right)
- {2 {\sf P} \over 3} \left[ \lambda_1 {\bf X} \times
\left( {\bf p}  + {{\sf P} \over 3}  {\bf X} \times {\bf X}' 
\right)\right]' \nonumber\\ \fl
&-& ( 2 \sigma^2 \lambda_1 {\bf X}' 
+ \lambda_2 {\bf p})'
\,.
\label{eq:hs2} 
\end{eqnarray}
These two couple first order equations reproduce 
the equilibrium condition ${\bf E}
= E {\bf n} = 0$, where $E$ is given by the l.h.s. of the 
Laplace-Young equation (\ref{eq:LY}), that follows from extremization of the 
energy. The first equation identifies both the momentum ${\bf 
p}$  and the Lagrange multipliers $\lambda_1, \lambda_2$. 
Using the constraints, one finds that the Lagrange multipliers
are proportional to the lapse and shift functions appearing in 
(\ref{eq:velo}):
\begin{equation}
\lambda_1 = {N \over 2 \sigma \sqrt{h}}\,,
\quad \quad
\lambda_2 = N_\parallel\,.
\label{eq:sbl}\end{equation}
These expressions for the Lagrange multipliers indicate that, given the curve ${\cal C}$,
the surface we generate depends only on the normal ${\bf l}$: the specific 
linear combination of ${\bf l}$ and ${\bf t}$ entering  
the velocity $\dot{\bf X}$ is gauge --
it is not determined by the Hamilton equations.
For the soap bubble, reparametrization invariance
manifests itself in a freedom within the velocity $\dot{\bf X}$ (for a fluid membrane 
this will become a freedom within the acceleration $\ddot{\bf X}$).
The definition (\ref{eq:p1}) of the momentum is reproduced when we substitute  
(\ref{eq:sbl}) for the Lagrange multipliers back into 
(\ref{eq:hs1}). Using this information
in the second of Hamilton's equations (\ref{eq:hs2}), one finds 
the following relationship:
\begin{equation}
- \sqrt{g} \; E \; {\bf n} = \dot{\bf p} + {\delta H \over \delta {\bf X}}\,,
\end{equation}
where $E$ is given by the left hand side of (\ref{eq:LY}). To check this,
one can use the geometrical expressions given in Appendix A.

Hamilton's equations 
(\ref{eq:hs1}), (\ref{eq:hs2}), together with the 
constraints (\ref{eq:sc1}), (\ref{eq:sc2}), provide an
alternative formulation of the single Euler-Lagrange equilibrium equation 
(\ref{eq:LY}). Once we have confirmed 
that this set of equations is equivalent to (\ref{eq:LY}), we 
can forget about the latter. 

The basic strategy to construct an equilibrium configuration is
to choose a closed curve in space defined by its position ${\bf X}(u,0)$.
We choose a momentum ${\bf p}(u,0)$
consistent with the constraints (\ref{eq:sc1}) and (\ref{eq:sc2}).
This corresponds to a direction.
Now we let ${\bf X}, {\bf p}$ evolve according to the 
Hamilton's equations (\ref{eq:hs1}) and
(\ref{eq:hs2})
for any $\lambda_1$ and $\lambda_2$ of our choosing.  
An equilibrium configuration ${\bf X}(u,t)$ will be generated.

In practice, when we discretize these equations, we will find that 
the canonical variables will stray off the constrained surface in 
phase space; the formalism itself suggests a strategy to 
 nudge them back.
One simply introduces 
 explicit damping terms proportional to the numerical value of the constraints
on the r.h.s. of Hamilton's equations.
It is also possible to tune $\lambda_1$ and $\lambda_2$ so as
pace the evolution most appropriately.

Unless we have been extremely lucky, we will find that the initial data we have prescribed generates 
a surface that does not close smoothly. This is not surprising:
after all, the unique non-self-intersecting geometry is a sphere of radius $R_0= 2\sigma/{\sf P}$. One needs 
to tune the initial conditions to produce this sphere. 

\section{ Phase space for the fluid membrane}

We turn now to the Hamiltonian formulation of the evolution of a curve
described by the action (\ref{eq:action}).
We first identify the phase space variables. Because the Lagrangian 
(\ref{eq:lagrange}) depends on the acceleration $\ddot{\bf X}$,
in addition to the position of the curve ${\bf X}$ and its conjugate 
momentum ${\bf p}$, the phase space will need to be enlarged 
to include
the velocity $\dot{\bf X}$ and its conjugate  momentum ${\bf P}$. 
(A pedagogical introduction to the Hamiltonian formulation of higher order
systems is given in the Appendix of \cite{hamaxi}.) 
The momenta are defined by the functional derivatives
\begin{eqnarray}
{\bf P} &=& {\delta L \over \delta \ddot{\bf X}}\,, \\
{\bf p} &=& {\delta L \over \delta \dot{\bf X}} -
\partial_t \left( {\delta L \over \delta \ddot{\bf X}} \right)\,.
\end{eqnarray}
Using the expression (\ref{eq:kpro}) for the mean curvature $K$, 
one immediately obtains for the momentum ${\bf P}$ conjugate to $\dot{\bf X}$, 
\begin{equation}
{\bf P} = - { \sqrt{h} \over N} \; (\kappa K  + \beta) \; {\bf n} \,.
\label{eq:hm}
\end{equation}
${\bf P}$ is normal to the surface, and transforms as a density under
a reparametrization of the curve. The part due to bending is proportional
to the mean curvature. 
We anticipate that this expression will be the content of one 
of Hamilton's equations for this system (see
(\ref{eq:h2}) below). Note that ${\bf P}$ 
 depends on ${\bf X}$, $\dot{\bf 
X}$, and $\ddot{\bf X}$.

The functional derivatives involved in the determination of the
momentum ${\bf p}$ are somewhat less simple.
A more convenient approach, generalizing that used in the
previous section for the soap bubble, see (\ref{eq:firsts}), focuses on 
the first
variation of the  energy. For  a system that depends on  the
acceleration, (\ref{eq:firsts}) generalizes to
\begin{equation}
\delta F [{\bf X}] = 
\int dA \; {\bf E}  \cdot \delta {\bf X}
 + \oint du \left({\bf p} \cdot \delta {\bf X} + {\bf P} \cdot \delta 
\dot{\bf X} \right)\,,
\label{eq:first}
\end{equation}
where ${\bf E}$  is the Euler-Lagrange derivative. Therefore, just as we 
did for a soap bubble, it is possible to read off the
momenta from this expression. To carry out the variation,
we  use variational results  described {\it e.g.} in \cite{Second},
such as (\ref{eq:va}), (\ref{eq:vv}), together with
\begin{equation}
\delta K = - {\bf n} \cdot \nabla^2 \delta {\bf X} + 2 K^{ab} {\bf e}_a
\cdot \partial_b \delta {\bf X}\,.
\end{equation}
The Euler-Lagrange derivative is normal, 
${\bf E} = E {\bf n}$, where $E$ is given by the l.h.s. of the
shape equation (\ref{eq:shape}). Let us focus our attention on the
second term of (\ref{eq:first}). Integrating by parts 
and using Stokes' theorem (\ref{eq:stokes}), one finds that
\begin{eqnarray}
\fl
 \oint du \left({\bf p} \cdot \delta {\bf X} + {\bf P} \cdot \delta 
\dot{\bf X} \right)
&=&  \oint du \sqrt{h} l_a \Big\{ \kappa \left[ (\nabla^a K) {\bf n} 
+ ({1 \over 2} K^2 g^{ab} - K K^{ab} )
{\bf e}_b \right] \nonumber \\ \fl
&+& \beta (K g^{ab} - K^{ab} )
{\bf e}_b  + \sigma {\bf e}^a - {{\sf P} \over 3} [ ({\bf n} \cdot {\bf 
X} ) {\bf e}^a - ( {\bf e}^a \cdot {\bf X} ) {\bf 
n} ] \Big\} \cdot \delta {\bf X} \nonumber \\ \fl
&-& \oint du \sqrt{h} l^a  ( \kappa K + \beta  ) {\bf n} \cdot \nabla_a 
\delta {\bf X} \,.
\label{eq:srf}
\end{eqnarray}
The terms in parenthesis spanning the first two lines 
contribute to the momentum ${\bf p}$.
The third line need some massaging to bring it into the form
(\ref{eq:first});
we use the definition of the velocity (\ref{eq:velo}), to decompose 
$ l^a \nabla_a \delta{\bf X} = 
N^{-1} ( \delta \dot{\bf X} 
- N_\parallel \delta {\bf X}' )$.
The third line of (\ref{eq:srf}) thus contributes to both ${\bf P}$ and ${\bf p}$:
\begin{eqnarray}
- \oint du \; \sqrt{h} ( \kappa K {\bf n} + \beta {\bf n} ) l^a \nabla_a 
\delta {\bf X} &=& 
\oint du \; {\bf P}
\cdot
\left( \delta \dot{\bf X} 
+ N_\parallel (\delta {\bf X}){}' \right)
\nonumber \\
&=& \oint du \; [ {\bf P}
\cdot \delta \dot{\bf X} 
- ( N_\parallel {\bf P} ){}' \cdot \delta {\bf X}]\,.
\end{eqnarray}
The definition (\ref{eq:hm})
of the momentum ${\bf P}$ is recovered, and we identify 
the momentum ${\bf p}$ conjugate to ${\bf X}$ as 
\begin{eqnarray}
\fl
{\bf p} &=& \sqrt{h} l_a \left\{   \kappa \left[ 
\left({1 \over 2} K^2 g^{ab} - K K^{ab} \right)
{\bf e}_b + (\nabla^a K) {\bf n}  
\right]
+ \beta  (K g^{ab} - K^{ab} )
{\bf e}_b \right\} \nonumber \\ \fl
&+& \sigma \sqrt{h} {\bf l}  - {{\sf P} \over 3} {\bf X} \times {\bf X}' 
+ ( N_\parallel {\bf P}){}'
\,. 
\label{eq:sm}
\end{eqnarray}
The momentum ${\bf p}$ also tranform as a density under reparametrization of 
the 
curve. 
Note that it depends on ${\bf X}$, $\dot{\bf X}$, $\ddot{\bf X}$ and 
${\bf X}$ with three dots.
For $\kappa = \beta = 0$ the momentum for a soap bubble (\ref{eq:p1}) is recovered.
We also anticipate that one  of Hamilton's 
equations will reproduce this expression (see (\ref{eq:ham3}) below). 
Geometrically, 
the momentum ${\bf p}$ given by (\ref{eq:sm}) is very simple;
${\bf p}$ is also related in a very direct way to the conserved stress 
for fluid membranes
derived by two of us in \cite{stress}. To illustrate the 
relationship, it is convenient 
to set ${\sf P}$ to vanish. The conserved stress for the energy (\ref{eq:model})
is
\begin{equation}
\fl
{\bf f}^a = \kappa \left[ \left( K K^{ab} - {1\over 2} K^2 g^{ab} \right)
{\bf e}_b - (\nabla^a K ) {\bf n} \right] + \beta \left(
K^{ab} - K g^{ab} \right) {\bf e}_b
- \sigma {\bf e}^a\,.
\end{equation}
Its divergence gives the shape equation as a conservation law:
\begin{equation}
\nabla_a {\bf f}^a = E {\bf n} = 0\,,
\label{eq:divo}
\end{equation}
where $E$ is given by the l.h.s. of (\ref{eq:shape}). 
We exploit the completeness of the surface tangent vectors 
$g^{ab}= l^a l^b +  t^a t^b$,
where $l_a = {\bf l}\cdot {\bf e}_a$, and $t_a = {\bf t}\cdot {\bf e}_a$ to
decompose the stress as
\begin{equation}
{\bf f}^a = {\bf f} \; l^a + {\bf g} \; t^a\,.
\label{eq:decs}
\end{equation} 
The canonical 
momentum ${\bf p}$ is given by (minus) the projection of the stress
along the normal to the curve $l^a$, up to a total divergence:
\begin{equation}
{\bf p} = - \sqrt{h} {\bf f} + ( N_\parallel {\bf P}){}'\,.
\label{eq:cs}
\end{equation}
Note that integrated over the closed curve:
 \begin{equation}
\oint du \, {\bf p} = -\oint du  \sqrt{h} l_a {\bf f}^a \,.
\label{eq:pint}
\end{equation}
The integrated momentum is the total force acting on the region enclosed by
the curve; if there are no external forces acting and ${\sf P}=0$, this integral will vanish.

One of Hamilton's equations will be an equation for $\dot {\bf p}$ 
(see (\ref{eq:ham4}) below).
Modulo the remaining canonical equations, this equation 
should reproduce the conservation law. 
Using the decomposition of the
stress tensor (\ref{eq:decs}) in (\ref{eq:divo}), we find
\begin{equation}
N \sqrt{h} \nabla_a {\bf f}^a 
= \partial_t (\sqrt{h} {\bf f} ) + ( N  {\bf g} 
- N_\parallel \sqrt{h} {\bf f}){}'\,.
\end{equation}
where we also make use of the identities (\ref{eq:a1}), (\ref{eq:a2}) given in  Appendix A. 
If we now use (\ref{eq:cs}), we find that the (densitized) shape equation 
can be cast as a continuity equation,
\begin{equation}
\sqrt{g} E {\bf n} = \sqrt{g} \nabla_a {\bf f}^a =
- \dot{\bf p} + {\bf m}{}' = 0\,,
\label{eq:conti}
\end{equation}
with a current ${\bf m}$ given by
\begin{equation}
{\bf m} = \partial_t ( N_\parallel {\bf P}) + N  {\bf g}  
+ N_\parallel {\bf p} - 
N_\parallel ( N_\parallel {\bf P}){}'\,. 
\label{eq:current}
\end{equation}
If now we
restore a non vanishing pressure {\sf P}, the continuity equation
(\ref{eq:conti}) acquires a source term, see (\ref{eq:ham4}) below.

\section{ Hamiltonian and constraints}

The canonical Hamiltonian is given by a Legendre transformation 
of the Lagrangian  $L [{\bf X}, \dot{\bf X} , \ddot{\bf X} ]$, as
given by
(\ref{eq:lagrange}), with
respect to both the velocity $\dot{\bf X}$ and the
acceleration $\ddot{\bf X}$,
\begin{equation}
H_0 [{\bf X}, {\bf p}; \dot{\bf X}, {\bf P} ]
= \oint du \; \left( 
{\bf P} \cdot \ddot{\bf X} + {\bf p} \cdot \dot{\bf X}  \right)
- L [{\bf X}, \dot{\bf X} , \ddot{\bf X} ]\,,
\end{equation}
where the definition of the momentum
${\bf P}$ is used to express the acceleration
$\ddot{\bf X}$ in terms of  the phase space variables. 
It should be emphasized that the term $ {\bf p} \cdot \dot{\bf X}  $ 
is left alone; the only place ${\bf p}$
appears in $H_0$ is in the first term.
The only tricky bit is the elimination of $\ddot{\bf X}$.
To do this, first note that
from the definition of ${\bf P}$ (\ref{eq:hm}), and 
the decomposition of the mean curvature (\ref{eq:kpro}), we have   
\begin{equation}
{\bf P} \cdot \ddot{\bf X}
=
{\kappa \sqrt{h} \over N^3 }
( {\bf n} \cdot \ddot{\bf X} )^2
- {\sqrt{h} \over N^3} (\kappa J + \beta N^2
) ( {\bf n} \cdot \ddot{\bf X} )\,;
\label{eq:leg}
\end{equation}
the quantity $J$, defined in (\ref{eq:J}), does
not depend on the acceleration. 
Setting 
\begin{equation}
L [{\bf X}, \dot{\bf X} , \ddot{\bf X} ] = 
\oint du \; {\cal L} ({\bf X}, \dot{\bf X} , \ddot{\bf X} )\,,
\end{equation}
we find that 
\begin{equation}
\fl {\bf P} \cdot \ddot{\bf X}
- {\cal L} ({\bf X}, \dot{\bf X} , \ddot{\bf X} )
= \sqrt{h} N \left[ {\kappa  \over 2 N^4 }   
( {\bf n} \cdot \ddot{\bf X} )^2
-{\kappa \over 2 N^4}   J^2 - {\beta  \over N^2 } J
- \sigma
+  {{\sf P}\over 3}   {\bf n} \cdot {\bf X} \right]\,;
\label{eq:leg1}
\end{equation}
the acceleration has been isolated in a single term.
We can express this term with respect to the phase space variables by 
inverting  the definition for $( {\bf n} \cdot \ddot{\bf X} )^2$:
\begin{equation}
{\bf P}
=
{\sqrt{h} \over N^3 }\left[\kappa 
( {\bf n} \cdot \ddot{\bf X} )
-  (\kappa J + \beta N^2)
\right] \,   {\bf n} 
\label{eq:leg2}
\end{equation}
implies 
\begin{equation}
( {\bf n} \cdot \ddot{\bf X} )^2
= {N^6 \over \kappa^2 h}
\left[ {\bf P} + {\sqrt{h} \over N} \left({\kappa \over N^2} J + \beta 
\right) {\bf
n}\right]^2\,.
\end{equation}
Substituting this this expression into (\ref{eq:leg1}), we identify  
the canonical Hamiltonian as
\begin{eqnarray}
\fl
H_0 [ {\bf X}, {\bf p}; \dot{\bf X}, {\bf P} ]
= \oint du \;  [{\bf p} \cdot \dot{\bf X} &+& 
{N^3 \over 2 \kappa \sqrt{h}} \, {\bf P}^2      
+ \left( J + N^2 {\beta \over \kappa} \right) \; {\bf P} \cdot {\bf n}
+ {N \sqrt{h} \beta^2 \over 2\kappa}  \nonumber \\ \fl
&-&  
\sigma N \sqrt{h}
+  {{\sf P}\over 3} N \sqrt{h}  ({\bf n} \cdot {\bf X}) ] \,.
\label{eq:hcan}
\end{eqnarray}
Unlike its soap bubble counterpart, this Hamiltonian does not vanish at the level of 
its definition, although it turns out that it does vanish anyway, see below.
It is quadratic in the higher momentum ${\bf P}$; it is linear in 
${\bf p}$ --- indeed, ${\bf p}$ only appears in the first term. 
The dependence on $\dot {\bf X}$ is rather more complicated: linearly 
in the first term
through ${\bf p}\cdot \dot {\bf X}$, and non-linearly everywhere else 
through its appearance in $N$, ${\bf n}$ and $J$. 
The dependence
on ${\bf X}$ is similarly complicated; 
it appears always through
its spatial derivatives with respect to the parameter $u$ along
the curve ${\cal C}$ with the  exception of the last term proportional
 to the pressure ${\sf P}$ in which ${\bf X}$ appears directly. 
While this dependence
does not break translation invariance, it signals that $\dot {\bf p}$ 
will not be a divergence.

The reparametrization invariance of the energy
manifests itself, at a Hamiltonian level, in the presence
of constraints: the phase space variables are not all independent
at a fixed value of the parameter $t$. 
Primary constraints appear because it is impossible to invert for the higher
momentum ${\bf P}$ in terms of the acceleration $\ddot{\bf X}$, just as, in 
Sect. 4, it was impossible to invert for the momentum ${\bf p}$ in terms
of the velocity $\dot{\bf X}$. Note, however, that this did not hinder the 
completion of the Legendre transformation.
The primary constraints are identified by the null eigenvectors of the
Hessian of the Lagrangian with respect to the higher derivatives
\begin{equation} 
{\cal H}_{ij} = {\delta^2 L [{\bf X}, \dot{\bf X}, 
\ddot{\bf X}]
\over \delta \ddot{ X}^i \delta \ddot{X}^j } = - \kappa \; {\sqrt{h} \over N^3} \; n_i n_j\,.
\end{equation}
There are two null eigenvectors, $\dot{\bf X}$ and ${\bf X}'$; we therefore
identify the two primary constraints
\begin{eqnarray}
C_1 &=& {\bf P} \cdot \dot{\bf X} = 0\,,
\label{eq:c1}
\\
C_2 &=& {\bf P} \cdot {\bf X}' = 0\,.
\label{eq:c2}
\end{eqnarray}
Their geometrical content is simple: the  
momentum ${\bf P}$ is normal to the surface.

The Hamiltonian 
function that generates 
the motion is given by adding the primary constraints to the canonical
Hamiltonian (\ref{eq:hcan}), giving the total Hamiltonian
\begin{equation}
H [{\bf X}, {\bf p}; \dot{\bf X}, {\bf P} ]
= H_0 [{\bf X}, {\bf p}; \dot{\bf X}, {\bf P} ]
 + \oint du \left( \lambda_1  C_1 + \lambda_2  C_2 \right)\,,
\label{eq:Htot}
\end{equation}
where $\lambda_1 = \lambda_1 (u,t), \lambda_2 = \lambda_2 (u,t)$ are 
arbitrary Lagrange multipliers that enforce the constraints. 

The Poisson bracket appropriate for this higher derivative system is,
for any two functions on the phase space  $f,g$,
\begin{equation}
\{ f , g \} = \oint du \; \left[ 
{\delta f \over \delta \dot{\bf X}} \cdot
{\delta g \over \delta {\bf P}}
+ {\delta f \over \delta {\bf X}} \cdot
{\delta g \over \delta {\bf p}}
- (f\leftrightarrow g ) \right]\,.
\label{eq:poisson}
\end{equation}

The time derivative of any function on phase space $f$ is then given by
its Poisson bracket with the total Hamiltonian (\ref{eq:Htot})
\begin{equation}
\dot{f} = \{ f , H \}\,.
\end{equation}

We have identified the primary constraints on the phase space variables.
For consistency, these constraints must be conserved in time. 
This requires us to introduce new
 {\it secondary} constraints. 
First, note that the primary constraints are in involution under the
Poisson bracket, {\it i.e.}  $\{ C_1 , C_2 \} = C_2 $.
The conservation in time of the primary constraints (\ref{eq:c1}) and 
(\ref{eq:c2}) implies the existence of  the secondary constraints, 
with $H_0 = \oint du \; {\cal H}_0$,
\begin{eqnarray}
S_1 &=& {\cal H}_0  = 0\,,
\label{eq:s1}
\\
S_2 &=& {\bf p} \cdot {\bf X}' + {\bf P} \cdot \dot {\bf X}' = 0\,.
\label{eq:s2}
\end{eqnarray}
A  detailed proof that these constraints follow from the conservation in time
of the primary constraints is provided in Appendix B. As we will show below, a far simpler derivation follows directly from reparametrization invariance. 

Among the secondary constriants 
is the vanishing of the canonical Hamiltonian, the
hallmark of reparametrization invariance. 
Together, the two secondary constraints completely fix the two tangential components of the momentum
${\bf p}$, ${\bf p}\cdot \dot{\bf X}$ and ${\bf p}\cdot {\bf X}'$. There are no 
other, tertiary constraints -- the four constraints we 
have identified are in involution under the Poisson brackets.

The secondary constraints are the generators of reparameterizations. 
In particular, $S_2$ generates reparametrizations tangential
to the curve ${\cal C}$, as follows from the fact that if we denote
by ${\bf Z}$ any of the phase space variables,
\begin{equation}
\delta {\bf Z} = \{ {\bf Z} , S_2 \} = {\bf Z}'\,.
\end{equation}
On the other hand, 
$S_1 = - {\cal H}_0$ 
generates reparametrizations off the curve ${\cal C}$, as it will
become clear in the next section, where we consider the evolution it
determines. The details of the model will enter through ${\cal H}_0$.

Conversely, one can show that 
reparametrization invariance implies all four counstraints:
the primary constraints (\ref{eq:c1}) and (\ref{eq:c2}) as well as the secondary constraints
(\ref{eq:s1}), (\ref{eq:s2}). To see this, let us re-examine the 
expression (\ref{eq:first}) for the first variation of the energy.
Consider first a temporal reparametrization $\delta {\bf X} = \omega (u,t) \dot{\bf X}$,
where the parameter $\omega$ is some function of the parameters $u$ and $t$.
In equilibrium, we have for the right hand side of (\ref{eq:first}) 
\begin{equation}
\oint du \left({\bf p} \cdot \delta {\bf X} + {\bf P} \cdot \delta 
\dot{\bf X} \right) =
\oint du \,\omega \left({\bf p} \cdot \dot{\bf X} + {\bf P} \cdot 
\ddot{\bf X} \right) + \oint du \,\dot \omega 
\, {\bf P} \cdot 
\dot{\bf X} 
\,,
\end{equation}
On the other hand, for
the left hand side of (\ref{eq:first}), we have
\begin{equation}
\delta F [{\bf X}] = \oint du \;\omega  L\,. 
\end{equation}
We conclude that 
\begin{equation}
 \oint du \;\omega  {\cal H}_0 + \dot \omega C_1 =0 \,. 
\end{equation}
Because $\omega$ and $\dot\omega$ are completely arbitrary,
we conclude that the invariance of $F$ under a redefinition of time
implies the constraints (\ref{eq:c1}) and (\ref{eq:s1}).
In exactly the same way,  
reparametrization invariance tangential to the curve implies the
constraints (\ref{eq:c2}) and (\ref{eq:s2}).
The stucture of the constraints depends only on the 
number of derivatives; it is independent of the details of the model.

This analysis also clarifies the role of the primary constraints.
In the transformation for the 
velocity induced by $\delta {\bf X} = \omega \dot{\bf X}$,
$\delta \dot {\bf X} = \omega \ddot{\bf X} + \dot \omega \dot {\bf X}$,
the second term is a scaling of $\dot {\bf X}$. 
The constraint $C_1$ is the generator of 
this scale transformation. An equal and opposite scaling of 
${\bf P}$ is required to preserve its Poisson bracket with $\dot{\bf X}$.

A result of reparametrization invariance is that the phase space variables provide
a redundant description of our system. The four constraints do, however, 
permit us to identify the independent degrees of freedom: these
are position of the curve ${\bf X}$ and its  
velocity $\dot{\bf X}$ as well as the normal components of 
${\bf P}$ and ${\bf p}$ (or, equivalently, $\ddot{\bf X}$ and ${\bf X}^{(3)}$ (with three dots)). 
Whereas both ${\bf X}$ and $\dot{\bf X}$ can be freely specified, 
the primary constraints kill the tangential components of 
${\bf P}$ and the secondary constraints fix the tangential components of ${\bf p}$ in terms of the 
remaining canonical variables. 

\section{ Hamilton's equations}

In this section, we present the complete set of Hamilton's equations that 
follow from the
Hamiltonian (\ref{eq:Htot}). 
We find it convenient to consider the 
equations in this order: \begin{equation}
\partial_t {\bf X} = {\delta H \over \delta {\bf p}}\,;\quad
\partial_t \dot{\bf X} = {\delta H \over \delta {\bf P}}\,;\quad
\partial_t {\bf P} = - {\delta H \over \delta \dot{\bf X}}\,;\quad
\partial_t {\bf p} = - {\delta H \over \delta {\bf X}}\,.
\end{equation}
There is one equation for each 
phase space variable; the equations are 
first order in time.

\vskip.25cm

\noindent (1) The 
first equation is 
\begin{equation}
\partial_t {\bf X}  = \delta H / \delta {\bf p} =
\dot{\bf X}\,.
\label{eq:h1}
\end{equation}
Note that the only dependence on ${\bf p}$ in the Hamiltonian is 
through the linear term ${\bf p} \cdot \dot{\bf X}$ appearing 
in the canonical Hamiltonian $H_0$ 
(\ref{eq:hcan}). Although it may appear to be a mere identity, its 
content is the identification of  the 
canonical variable $\dot {\bf X}$ with the time derivative of ${\bf X}$.
  
\vskip.25cm

\noindent (2) The second of Hamilton's equation 
$
\partial_t \dot{\bf X} = \delta H /\delta {\bf P}$ 
is (see Appendix B):
\begin{equation}
\partial_t \dot{\bf X} = \ddot{\bf X} = 
{N^3 \over  \kappa \sqrt{h}} {\bf P} 
+ \left( J + N^2 {\beta \over \kappa} \right) {\bf n}
+ \lambda_1 \dot{\bf X} + \lambda_2 {\bf X}'\,.
\label{eq:h2}
\end{equation}
Note that the r.h.s. does not depend on the momentum ${\bf p}$ conjugate 
to ${\bf X}$.
The content of this equation is to identify the momentum 
${\bf P}$ in terms of $\ddot{\bf X}, \dot{\bf X}$ and ${\bf X}$, 
reproducing the expression (\ref{eq:hm}). This equation also identifies
the form of the Lagrange multipliers $\lambda_1$ and $\lambda_2$. As we 
show in the next section, they are proportional to the tangential
components of the acceleration $\ddot{\bf X}$.

\vskip.25cm

\noindent (3) For the third of Hamilton's equation 
$\partial_t {\bf P} = - \delta H /\delta \dot{\bf X}$,
we obtain (see Appendix C):
\begin{eqnarray}
\fl
\partial_t {\bf P} 
&=&  
- {\bf p} + 2 h^{-1}({\bf p}\cdot {\bf X'}) {\bf X}'
+ 2 ( N_\parallel {\bf P}){}' 
+ \left\{ h^{-1} N^{-1}
 \left[ 2 N_\parallel h ({\bf 
l}\cdot \dot{\bf X}') 
- \dot{\bf X}^2  \; ({\bf l}\cdot {\bf X}'')\right] - \lambda_1 \right\} \;
{\bf P} 
\nonumber \\ \fl
&+&   
 2h^{-1} ({\bf P}\cdot {\bf X}'') \; \dot{\bf X}
- \left[ {3N^2 \over  2 \kappa \sqrt{h}} {\bf P}^2  
+ {2N\beta \over \kappa} ({\bf P}\cdot {\bf n})
+ {\beta^2 \sqrt{h}\over 2\kappa}
- \sigma \,\sqrt{h} \right] \; {\bf l} 
- {{\sf P} \over 3}  {\bf X} \times {\bf X}'
\,.
\label{eq:ham3}
\end{eqnarray} 
This equation identifies the 
momentum ${\bf p}$ in terms of 
${\bf X},\dot{\bf X}$, ${\bf P}$ and $\dot{\bf P}$.
It is independent of $\lambda_2$.
This presentation of ${\bf p}$ is very different from the 
form written down in (\ref{eq:sm}) where it is cast as
a functional of 
${\bf X},\dot{\bf X},\ddot{\bf X}$ and ${\bf X}$ with three dots. 
A little work is required 
to establish that indeed the two are the same ${\bf p}$. 
We will return to this issue in the 
next section.
 
At this point, it is worth emphasizing that the secondary 
constraints identified
in the previous section, (\ref{eq:s1}) and (\ref{eq:s2}), 
specify the tangential part of ${\bf p}$ in terms 
of the remaining phase space variables, and initial data
will need to be chosen accordingly.

\vskip.25cm
\noindent (4) Finally, the fourth Hamilton equation 
$\partial_t {\bf p} = - \delta H /\delta {\bf X}$
is
\begin{equation}
\partial_t {\bf p} = \dot{\bf p} 
= - {\bf m}' - {{\sf P}  \over 3} N 
\sqrt{h} {\bf n}\,,
\label{eq:ham4}
\end{equation}
where the current ${\bf m}$ is given by
\begin{eqnarray} \fl
{\bf m} &=&  
\left[ {3N^2\over 2\kappa \sqrt{h}}N_\parallel
{\bf P}^2 + {2\beta \over \kappa} N N_\parallel
({\bf P}\cdot {\bf n}) - 2Nh^{-1} ({\bf P}\cdot  \dot{\bf X}')
+ \sqrt{h} N_\parallel \left( {\beta^2 \over 
2\kappa} - \sigma 
\right)  \right]{\bf l} \nonumber \\ \fl
&+&  \left[ h^{-3/2}
{N^3 \over 2\kappa }
{\bf P}^2 - h^{-2}\dot{\bf X}^2 ({\bf P}\cdot
{\bf X}'')  + 2 h^{-1} N_\parallel ({\bf P} \cdot \dot{\bf X}') 
- h^{-1/2} N \left( {\beta^2 \over 2\kappa}
- \sigma \right)
\right]\,{\bf X}' 
     \nonumber \\ \fl
&+&  {1 \over N \sqrt{h}} \left[
h^{-1}  \dot{\bf X}^2 ({\bf n} \times \dot{\bf X}\cdot {\bf X}'')
- 2 N_\parallel ({\bf n} \times \dot{\bf X} \cdot \dot{\bf X}')
\right] \,{\bf P} -  \left( \dot{\bf X}^2 h^{-1}{\bf P}\right){}'
\nonumber \\ \fl
&-& \lambda_2 \,{\bf P} + {{\sf P} \over 3}  {\bf X} 
\times \dot{\bf X}\,.  
\label{eq:curro}
\end{eqnarray}
Modulo the other Hamilton's equations, which establish the relationships
between the dynamical variables appearing on the r.h.s., 
this is  the shape equation (\ref{eq:shape}) cast in canonical form.
We will show this explicitly in the next section.
Note that (\ref{eq:ham4}) assumes the form of a continuity equation. 
The origin of the source term is the 
explicit dependence of the Hamiltonian $H$ on
${\bf X}$ appearing in the volume when the pressure ${\sf P}$ is non-vanishing.

To summarize, Hamiltons equations (\ref{eq:h1}), (\ref{eq:h2}), 
(\ref{eq:ham3}) and (\ref{eq:ham4}) together 
with the primary constraints (\ref{eq:c1}), (\ref{eq:c2}), and the
secondary constraints
(\ref{eq:s1}), (\ref{eq:s2}), 
provide an alternative formulation of the shape equation
(\ref{eq:shape}). 
Just as it does for the Einstein equations in General Relativity, this
formulation  calls for a change of perspective. We started with  
with a surface satisfying 
the shape equation. There is now no
surface, only a curve. 

The recipe for reconstructing a surface is simple:
($i$) We choose any closed curve in space  (${\bf X}(0,u)$). 
($ii$) We  are free to specify its velocity $\dot {\bf X} (0,u) $. At this point we have
defined a basis in space adapted to the curve given by 
the tangent to the curve,${\bf X}' (0,u)$, $\dot{\bf X} (0,u)$ and their 
cross product, proportional to the unit normal ${\bf n} (0,u)$. 
The remaining initial data on this curve must be consistent with the constraints.  
($iii$) To satisfy the primary constraints
(\ref{eq:c1}) and (\ref{eq:c2}),
the  momentum ${\bf P}$ is  chosen to be orthogonal both to $\dot{\bf X}$ 
and to the tangent to
the curve ${\bf X}'$.  
($iv$) The momentum 
${\bf p}$ is constrained by  
the secondary constraints (\ref{eq:s1}) and (\ref{eq:s2}) which determine its 
two tangential components 
in terms of the initial data we have already specified. 
(Note that ${\bf p}$ must also be consistent with the 
integrability condition associated with a closed curve given by 
(\ref{eq:pint})).
($v$) We choose appropriate values for the two
functions $\lambda_1$ and $\lambda_2$ --- this corresponds to a 
choice of coordinates. As we show in the next section, they are
proportional to the tangential components of $\ddot{\bf X}$. 
($vi$) We now let the system evolve according to Hamilton's equations. 
An equilibrium surface ${\bf X}(u,t)$ satisfying the shape equation 
will be generated.

\section{ Shape equation from Hamilton's equations with constraints}

In this section, we show how the Hamilton equations, together with
the constraints,  combine to reproduce the shape equation (\ref{eq:shape}).

\vskip .25cm

The second Hamilton equation (\ref{eq:h2})
identifies ${\bf P}$ and the form of the Lagrange multipliers.
We dot (\ref{eq:h2}) with 
the normal
${\bf n}$, and we use  the primary constraints 
(\ref{eq:c1}), 
(\ref{eq:c2}), to obtain
\begin{equation}
N^3 {\bf P} \cdot {\bf n} = \kappa \sqrt{h} ( \ddot{\bf X} \cdot 
{\bf n} - J) - N^2 \sqrt{h} \beta\,.
\end{equation}
As the primary constraints tell us that ${\bf P}$ is purely normal 
to the surface, using the expression (\ref{eq:kpro}) for the
mean curvature $K$, we recover the expression (\ref{eq:hm}) for
${\bf P}$.

To obtain the form of the Lagrange multipliers we dot  (\ref{eq:h2})
with $\dot{\bf X}$ and ${\bf X}'$, and use again the primary constraints
(\ref{eq:c1}), 
(\ref{eq:c2}), 
so 
that we find
\begin{eqnarray}
\lambda_1 &=& {\dot{\bf X} \cdot \ddot{\bf X} \over  
\dot{\bf X}^2}\,, 
\label{eq:lag1}\\
\lambda_2 &=& h^{-1} {\bf X}' \cdot \ddot{\bf X} - N_\parallel \lambda_1
= {N \over \sqrt{h} \dot{\bf X}^2} {\bf n} \times \dot{\bf X} \cdot 
\ddot{\bf X}\,.
\label{eq:lag2}
\end{eqnarray}
The first tells us  that $\lambda_1$ has the geometrical meaning of 
the 
affine connection along the integral curve of $\dot{\bf X}$;
the second that $\lambda_2$ is proportional to the tangential component
of the acceleration $\ddot{\bf X}$ orthogonal to the velocity $\dot{\bf X}$.
Therefore, for this higher derivative system, the components
of the acceleration tangential to the surface are arbitrary; they are
the gauge part of the evolution, whereas the normal component is 
proportional
to the momentum ${\bf P}$. In the case of a soap bubble,
the Lagrange multipliers were proportional to the components of the
velocity -- this is its natural higher order generalization.

\vskip .25cm

The third Hamilton equation (\ref{eq:ham3})
identifies the 
form of the momentum ${\bf p}$, as given by (\ref{eq:sm}).
To see this is not immediate. 
To facilitate the
comparison, we consider first (\ref{eq:sm}).
We use the Gauss-Weingarten equation (\ref{eq:gw2}), and the
expression (\ref{eq:hm}) for ${\bf P}$, to obtain for ${\bf p}$ 
\begin{eqnarray} \fl
{\bf p} &=& \sqrt{h} l_a \left[   \kappa 
\left({1 \over 2} K^2 g^{ab} - 2 K K^{ab} \right)
+ \beta  (K g^{ab} - 2 K^{ab} )
\right] \; {\bf e}_b  -\dot{\bf P} + 2 ( N_\parallel 
{\bf P}){}' \nonumber \\ \fl
&-& \left[ N_\parallel{}'  + {N^2 \over \sqrt{h}} l^a \nabla_a
\left( {\sqrt{h} \over N} \right) \right] {\bf P} 
+ \sigma \sqrt{h} \; {\bf l}  - {{\sf P} \over 3} {\bf X} \times {\bf X}' 
\,. 
\label{eq:sm2}
\end{eqnarray}
Now  using the completeness relation on the surface
$g^{ab} = l^a l^b + t^a t^b $, we have that
\begin{equation}
K^{ab} l_a {\bf e}_b = K {\bf l} + 
{1 \over h N} 
\left[ ({\bf n} \cdot 
{\bf X}'' ) \dot{\bf X} - ( {\bf n} \cdot \dot{\bf X}')  {\bf 
X}'\right]\,,
\label{eq:prk}
\end{equation}
and inserting this expression in (\ref{eq:sm2}) we obtain 
the 
alternative expression for ${\bf p}$
\begin{eqnarray} \fl
{\bf p} &=& \sqrt{h}  \kappa \left\{ 
 - {3 \over 2} K^2 \; {\bf l}  - 2  
h^{-1} N^{-1} K  \left[ ({\bf n} \cdot 
{\bf X}'' ) \dot{\bf X} - h^{-1} N^{-1} 
( {\bf n} \cdot \dot{\bf X}')  {\bf 
X}'\right] \right\} \nonumber \\ \fl &-& 
\beta  \left\{  \sqrt{h} K \; {\bf l} + 2
h^{-1/2} N^{-1} 
\left[ ({\bf n} \cdot 
{\bf X}'' ) \dot{\bf X} - ( {\bf n} \cdot \dot{\bf X}')  {\bf 
X}'\right] \right\}
-\dot{\bf P} + 2 ( N_\parallel 
{\bf P}){}' \nonumber \\ \fl
&-& \left[ N_\parallel{}'  + {N^2 \over \sqrt{h}} l^a \nabla_a
\left( {\sqrt{h} \over N} \right) \right] {\bf P} 
+ \sigma \sqrt{h} \; {\bf l}  - {{\sf P} \over 3} {\bf X} \times {\bf X}' 
\,. 
\label{eq:sm3}
\end{eqnarray}
Using  the information gathered from the second 
Hamilton equation (\ref{eq:h2}), {\it i.e.}
(\ref{eq:hm}) for the momentum ${\bf P}$, and (\ref{eq:lag1}) for the
Lagrange multiplier 
$\lambda_1$, it is a simple matter of algebra to see that
the third Hamilton equation (\ref{eq:ham3}) coincides with
the expression (\ref{eq:sm3}) for ${\bf p}$.

\vskip .25cm

Finally, the fourth Hamilton equation (\ref{eq:ham4}) is 
(\ref{eq:conti}), that we have derived in Sect. 4 from the
conservation of the stress tensor at equilibrium. To see this,
requires using the specific form ${\bf P}, \lambda_1, \lambda_2$
in the expression (\ref{eq:curro}) for the current ${\bf m}$, and
comparison with the right hand side of the alternative expression
(\ref{eq:current}).

\section{Concluding remarks}

In this paper, we have presented a Hamiltonian formulation
of the shape equation for arbitrary, not necessarily axisymmetric,
configurations of lipid vesicles. In this description the surface 
geometry is reconstructed from a closed curve. This is possible because 
the conserved internal stresses which underpin the equilibrium geometry 
are transmitted across closed curves; 
the shape equation is a conservation law for these stresses.

This formulation of the shape equation
offers a new approach for constructing equilibrium configurations
based on the numerical solution of Hamilton's equations.
Whether its implementation  turns out to be tractable, in practice,
remains to be seen. There is no problem 
generating equilibrium surfaces: any initial data consistent with the 
constraints will do this. The subtlety is identifying initial data 
that are consistent with a surface which closes 
smoothly. In general, the curve will self-intersect and
singularities will arise. The initial data will thus need
to be tuned accordingly. This problem is a global one.
Progress is likely to be limited until a second problem 
is tackled: what is the most appropriate choice of Lagrange multipliers
for any given initial data? This will  fix the parametrization of the surface. 
There may not be a single choice which is appropriate everywhere.


\ack
We thank Markus Deserno and Martin M\"uller for useful comments.
RC thanks the Aspen Center for Physics for 
hospitality where part of this work was carried out.
We received partial support from CONACyT grants 44974-F and C01-41639.

\vspace{1cm}

\noindent{\bf  APPENDIX A}

\vspace{.5cm}

In this Appendix we collect various geometrical formulae that 
are useful at intermediate
stages of the calculations that we have performed. 
The Frenet-Serret equations for the curve ${\cal C}$, seen as   
living on $\Sigma$ are, with $\nabla_s = t^a \nabla_a $:
\begin{eqnarray} 
\nabla_s t^a &=& k \; l^a \,, 
\\
\nabla_s l^a &=& -   k \; t^a\,,
\end{eqnarray}
where $k$ is the geodesic curvature for the curve.
The Frenet-Serret equations for the curve ${\cal C}$ seen as living in 
space are \begin{eqnarray}
\nabla_s {\bf t} &=&   k_1 {\bf l} + k_2 {\bf n} \\
\nabla_s {\bf l} &=&  - k_1  {\bf t} + k_3 {\bf n}\\
\nabla_s {\bf n}  &=&  - k_2  {\bf t} - k_3 {\bf l}
\end{eqnarray}
where $k_1 = k$ is the geodesic curvature, and $k_2, k_3$ are given by
the
projections of the surface extrinsic curvature by $k_2 = - K_{ab} t^a t^b$,
$k_3 = - K_{ab} t^a l^b$.

The evolution in the parameter $t$ of the basis $\{ {\bf X}' , {\bf l}, 
{\bf n} \}$ can be derived
from the fact that partial derivatives commute, so that $ \dot{\bf X'} = 
(\dot{\bf X}){}'$. It follows that
\begin{eqnarray}
\dot{\bf X}' &=& (\partial_u N + N_\parallel k ) \; {\bf l}
+ (\partial_u N_\parallel - h^{-1} N k ) \; {\bf X}'
+ ({\bf n} \cdot \dot{\bf X}' ) \; {\bf n}\,,
\\
\dot{\bf l} &=&
- h^{-1} ( N' + N_\parallel k ) \; {\bf X}'
+ N^{-1}[({\bf n} \cdot \ddot{\bf X}) - N_\parallel ({\bf n} \cdot
\dot{X}')] \; {\bf 
n}\,,  \\
\dot{\bf n} &=&  -h^{-1} 
 ({\bf n} \cdot \dot{\bf X}' )
\; {\bf X}' -
 N^{-1}[
({\bf n} \cdot \ddot{\bf X}) 
- N_\parallel ({\bf n} \cdot
\dot{X}')] 
 \; {\bf l}\,.
\end{eqnarray}
Note that
\begin{equation}
\partial_t \sqrt{h} = - \sqrt{h} N k + \sqrt{h}\partial_u N_\parallel\,.
\end{equation}

The acceleration is
\begin{equation}
\fl
\ddot{\bf X} = (\dot{N}_\parallel - h^{-1} N \partial_u N 
+ N_\parallel \partial_u N_\parallel
- 2 h^{-1} N N_\parallel k ) \; {\bf X}' +
(\dot{N} + N_\parallel \partial_u N + N_\parallel^2
k ) \; {\bf l}  - ({\bf n} \cdot \ddot{\bf X}) 
 \; {\bf n}\,.
\end{equation}

We also note the identities
\begin{eqnarray}
\nabla_a l^a &=& -  k\, 
\label{eq:a1}\\
 \nabla_a t^a &=& N^{-1} \nabla_s N\,.
\label{eq:a2}
\end{eqnarray}

\vspace{1cm}

\noindent{\bf  APPENDIX B}

\vspace{.5cm}

In this Appendix, we show how the secondary constraints (\ref{eq:s1}),
(\ref{eq:s2}) follow from the conservation in time of the primary
constraints (\ref{eq:c1}), (\ref{eq:c2}). 
The first thing to note is that the primary constraints are in involution
under the Poisson bracket (\ref{eq:poisson}),
\begin{equation}
\{ C_1 , C_2 \} = C_2 \approx 0\,,
\end{equation}
where the symbol $\approx$ denotes equality modulo the constraints 
themselves.

The time derivative of the first primary constraint (\ref{eq:c1}) is
\begin{equation}
\dot{C}_1 = \{ C_1 , H \} \approx \{ C_1 , H_0 \} = {\bf P} \cdot {\delta H_0 
\over \delta {\bf P}} - \dot{\bf X} \cdot {\delta H_0 \over \delta \dot{\bf X}}
\,.
\end{equation}
We need to evaluate the functional derivatives of the canonical Hamiltonian
functional (\ref{eq:hcan}).
For the functional derivative of the canonical Hamiltonian with respect
to ${\bf P}$, we have immediately that
\begin{equation}
{\delta H_0 \over \delta {\bf P}} = {N^3 \over  \kappa \sqrt{h}} {\bf P} 
+ \left( J + N^2 {\beta \over \kappa} \right) {\bf n}\,.
\end{equation}
In order to evaluate the functional derivative of $H_0$ with respect to 
the velocity $\dot{\bf X}$, we hold ${\bf X}$ fixed, and consider the 
variation with respect to $\dot{\bf X}$
of 
the geometric quantities that appear in the canonical Hamiltonian: 
\begin{eqnarray}
\delta N &=& {\bf l} \cdot \delta \dot{\bf X}
\,, \nonumber\\
\delta N_\parallel &=& h^{-1} \; {\bf X}' \cdot \delta \dot{\bf X}
\,, \nonumber\\
\delta {\bf n} &=& - N^{-1} \; ({\bf n} \cdot \delta \dot{\bf X}) \; {\bf 
l}\,.
\nonumber
 \end{eqnarray}
For the part of the mean curvature that does not depend on the
acceleration $\ddot{\bf X}$, defined as $J$ in ({\ref{eq:J}), it follows that
\begin{eqnarray}
\fl
\delta J &=& h^{-1} 
\left\{ 2 ({\bf n}\cdot \dot{\bf X}' )
{\bf X}'  +  N^{-1} \left[ \dot{\bf X}^2  
( {\bf l}\cdot {\bf X}'')
- 2 N_\parallel h ({\bf 
l}\cdot \dot{\bf X}') \right]
\; {\bf n}
- 2  ({\bf n}\cdot {\bf X}'')\dot{\bf X}
\right\}
\cdot \delta 
\dot{\bf X} \nonumber \\ \fl
&+& 2 N_\parallel \; {\bf n}\cdot (\delta \dot{\bf X}){}' \,.
\nonumber
\end{eqnarray}
Using these expressions we have
\begin{eqnarray}\fl
{\delta H_0 \over \delta \dot{\bf X}} &=&
{\bf p} + \left[ {3N^2 \over  2 \kappa \sqrt{h}} {\bf P}^2  
+ {2N\beta \over \kappa} ({\bf P}\cdot {\bf n})
+ {\beta^2 \sqrt{h}\over 2\kappa}
- \sigma \,\sqrt{h} \right] \; {\bf l} +
2 h^{-1}({\bf n} \cdot \dot{\bf X'}) 
({\bf P} \cdot {\bf n})\; 
{\bf X}' \nonumber \\ \fl
&-& 2 [ N_\parallel ({\bf P} \cdot {\bf n}) {\bf n} ]' 
+ h^{-1} N^{-1} ({\bf P} \cdot {\bf n})\; 
 \left[  \dot{\bf X}^2  \; ({\bf l}\cdot {\bf X}''
- 2 N_\parallel h ({\bf 
l}\cdot \dot{\bf X}') 
)\right] \;
{\bf n} \nonumber \\ \fl &-&
 2h^{-1} ({\bf n}\cdot {\bf X}'') ({\bf P} \cdot {\bf n})
\; \dot{\bf X}
+ {{\sf P} \over 3}  {\bf X} \times {\bf X}'\,.
\end{eqnarray}
We can simplify this expression using the completeness relation
$\delta^{ij} = n^i n^j + l^i l^j + h^{-1} \dot{X}^i \dot{X}^j$, so that, 
modulo the constraints, 
$({\bf P} \cdot {\bf n}) 
{\bf n} \approx {\bf P}$. Hence
\begin{eqnarray} \fl
{\delta H_0 \over \delta \dot{\bf X}} &\approx&
{\bf p} + \left[ {3N^2 \over  2 \kappa \sqrt{h}} {\bf P}^2  
+ {2N\beta \over \kappa} ({\bf P}\cdot {\bf n})
+ {\beta^2 \sqrt{h}\over 2\kappa}
- \sigma \,\sqrt{h} \right] \; {\bf l} +
2 h^{-1}({\bf P} \cdot \dot{\bf X'}) 
{\bf X}' - 2 ( N_\parallel {\bf P})' 
\nonumber \\ \fl
&+& h^{-1} N^{-1}
 \left[  \dot{\bf X}^2  \; ({\bf l}\cdot {\bf X}''
- 2 N_\parallel h ({\bf 
l}\cdot \dot{\bf X}') 
)\right] \;
{\bf P} -
 2h^{-1} ({\bf P}\cdot {\bf X}'') \; \dot{\bf X}
+ {{\sf P} \over 3}  {\bf X} \times {\bf X}'\,.
\label{eq:hdotx}
\end{eqnarray}
and one finds that the conservation in time of $C_1$ gives
the vanishing of the canonical Hamiltonian density ${\cal H}_0$,
defined by $ H_0 = \oint du \; {\cal H}_0$: 
\begin{equation}
\dot{C}_1 \approx - {\cal H}_0 = 0\,.
\end{equation}
To see that this expression holds it is essential to recognize that
\begin{equation}
J 
({\bf P} \cdot {\bf n}) 
\approx N_\parallel ({\bf P} \cdot \dot{\bf X}')
- (N_\parallel {\bf P})' \cdot \dot{\bf X} -  h^{-1} \dot{\bf X}^2 
({\bf P} \cdot {\bf X}'')\,,
\end{equation}
where again one uses the fact that $({\bf P} \cdot {\bf n}) 
{\bf n} \approx {\bf P}$, and $(N_\parallel {\bf P})' \cdot {\bf X}'
\approx - N_\parallel ({\bf P} \cdot \dot{\bf X}')$ . 

Similarly, the conservation in time of the second primary constraint 
(\ref{eq:c2}) gives the secondary constraint (\ref{eq:s2}):
\begin{equation}
\dot{C}_2 = \{ C_2 , H \} \approx \{ C_2 , H_0 \} = 
- {\bf X}' \cdot {\delta H_0 
\over \delta \dot{\bf X}} - \dot{\bf X} 
\cdot 
{\bf P}' \approx - S_2 = 0\,.
\end{equation}

\vspace{1cm}

\noindent{\bf APPENDIX C}

\vspace{.5cm}

In this Appendix we fill in the details of the derivations of the
third and fourth Hamilton equations (\ref{eq:ham3}) and (\ref{eq:ham4}).
Consider first the third Hamilton equation, 
$\dot{\bf P}= - \delta H / \delta \dot{\bf X}$. We use (\ref{eq:hdotx})
from Appendix B, and we obtain immediately
\begin{eqnarray} \fl
\dot{\bf P} &=& 
 - {\bf p} 
- \left[ {3N^2 \over  2 \kappa \sqrt{h}} {\bf P}^2  
+ {2N\beta \over \kappa} ({\bf P}\cdot {\bf n})
+ {\beta^2 \sqrt{h}\over 2\kappa}
- \sigma \,\sqrt{h} \right]
\; {\bf l} 
+ 2 \left( N_\parallel {\bf P} \right){}'
-  2 h^{-1}({\bf P}\cdot \dot{\bf X}')
{\bf X}'  
\nonumber \\ \fl
&+&  h^{-1} N^{-1} \left[ 2 N_\parallel h ({\bf 
l}\cdot  \dot{\bf X}') 
- \dot{\bf X}^2 ({\bf l}\cdot  {\bf X}'')\right]
{\bf P}
+ 2h^{-1} ({\bf P}\cdot {\bf X}'')\dot{\bf X}
\nonumber \\ \fl
&-&  {{\sf P} \over 3} {\bf X} \times {\bf X}'
- \lambda_1 \,{\bf P}\,.
\label{eq:APdot2}
\end{eqnarray} 
Using the secondary constraint (\ref{eq:s2}) in the last 
term of the first line
gives the third Hamilton equation in the form given in the 
text, (\ref{eq:ham3}).

Let us turn  to the derivation of the fourth Hamilton equation
$ \dot{\bf p} = - \delta H / \delta {\bf X}$. 
Now we hold $\dot{\bf X}$ fixed, and consider a variation of
of the geometric quantities
that appear in the canonical Hamiltonian w.r.t. ${\bf X}$:
\begin{eqnarray}
\delta {\bf X}' &=& 
(\delta {\bf X}){}' \,, \nonumber \\
\delta {\bf l}&=& -  N^{-1} N_\parallel [ {\bf n}\cdot 
(\delta {\bf X}){}'] \,{\bf n} -
h^{-1} [{\bf l}\cdot (\delta {\bf X}){}' ] \,{\bf X}'\,, \nonumber
\\
\delta {\bf n}&=&  \left( N^{-1} N_\parallel  {\bf l} - h^{-1} 
{\bf X}' \right)  [{\bf n}\cdot (\delta {\bf X}){}' ] =
N^{-1} h^{-1/2} ({\bf n} \times \dot{\bf X}) {\bf n} \cdot (\delta{\bf X})' 
\,, \nonumber \\
\delta N &=& - N_\parallel\, {\bf l}\cdot (\delta
{\bf X})'\,, \nonumber
\\
\delta N_\parallel &=& 
h^{-1}\, \left[ N {\bf l}\cdot (\delta {\bf X}){}'  - N_\parallel 
\,{\bf X}' \cdot (\delta {\bf X})' \right]\,. \nonumber
\end{eqnarray}
For the quantity $J$ this implies
\begin{eqnarray} \fl
\delta J &=& 2Nh^{-1}({\bf n}\cdot \dot{\bf X}')
{\bf l}\cdot (\delta {\bf X}){}'
+ 2 \left[ h^{-2}\dot{\bf X}^2 ({\bf n}\cdot
{\bf X}'')
 - N_\parallel h^{-1}({\bf n}\cdot \dot{\bf X}')\right]
{\bf X}' \cdot (\delta {\bf X}){}' \nonumber \\ \fl
&+& \left[2 {N_\parallel ^2 \over N}({\bf l}\cdot  
\dot{\bf X}')
- N^{-1} N_\parallel h^{-1}\dot{\bf X}^2({\bf l}\cdot 
{\bf X}'') - 2N_\parallel h^{-1}({\bf X}' \cdot \dot{\bf X}')
+ h^{-2} \dot{\bf X}^2 ({\bf X}' \cdot {\bf X}'')
\right] 
{\bf n}\cdot (\delta {\bf X}){}' 
\nonumber \\ \fl
&-& 
 h^{-1}\dot{\bf X}^2 \; {\bf n}\cdot 
(\delta {\bf X}){}''\,.
\end{eqnarray}
Using these expressions,  and exploiting again the fact that
$ ({\bf P} \cdot {\bf n}) {\bf n}  \approx {\bf P} $,
we find that the fourth Hamilton equation is
\begin{eqnarray} \fl
\dot{\bf p} &=& 
- \partial_u \left\lbrace
\left[ {3N^2\over 2\kappa \sqrt{h}}N_\parallel
{\bf P}^2 + {2\beta \over \kappa} N N_\parallel
({\bf P}\cdot {\bf n}) - 2Nh^{-1} ({\bf P}\cdot  \dot{\bf X}')
+ \sqrt{h} N_\parallel \left( {\beta^2 \over 
2\kappa} - \sigma 
\right)  \right]{\bf l} \right. \nonumber \\ \fl
&+& \left. \left[ h^{-3/2}
{N^3 \over 2\kappa }
{\bf P}^2 - h^{-2}\dot{\bf X}^2 ({\bf P}\cdot
{\bf X}'')  + 2 h^{-1} N_\parallel ({\bf P} \cdot {\bf X}') 
- h^{-1/2} N \left( {\beta^2 \over 2\kappa}
- \sigma \right)
\right]\,{\bf u} 
    \right. \nonumber \\ \fl
&+& \left. {1 \over N} \left[
h^{-1} N_\parallel \dot{\bf X}^2 ({\bf l}\cdot {\bf X}'')
- 2 N_\parallel^2 ({\bf l} \cdot \dot{\bf X}')
+ 2 N N_\parallel h^{-1} ({\bf X}' \cdot \dot{\bf X}') - h^{-2} N \dot{\bf 
X}^2 ({\bf X}' \cdot {\bf X}'' )
\right] \,{\bf P} \right. \nonumber \\ \fl
&-& \left. \left( \dot{\bf X}^2 h^{-1}{\bf P}\right)'
- \lambda_2 \,{\bf P} + {{\sf P} \over 3}  {\bf X} 
\times \dot{\bf X}  
\right\rbrace - {{\sf P} \over 3} 
N\sqrt{h}
\,{\bf n}\,.  
\end{eqnarray}
This gives the fourth Hamilton equation in the form given in 
the text, (\ref{eq:ham4}).

\end{document}